\documentclass[aps,prl,a4paper,tightenlines,amsmath,amssymb,twocolumn,floatfix,final]{revtex4}
\usepackage{latexsym} \usepackage{amssymb}
\usepackage[dvips]{graphics} \usepackage[pdftex]{graphicx}

 \newcommand{\beq}{\begin{eqnarray}}
\newcommand{\eeq}{\end{eqnarray}} \newcommand{\be}{\begin{eqnarray}}
\newcommand{\ee}{\end{eqnarray}}

\newcommand{\grad}{\ensuremath{^{\circ}}}
\usepackage[latin1]{inputenc} \usepackage{amsmath}
\usepackage{amsfonts} \usepackage{amssymb} \usepackage{amsthm}

\begin{document}
\title{Detecting Seasonal Changes in the Fundamental Constants}
\author{Douglas J. Shaw}
\email{D.Shaw@damtp.cam.ac.uk}
\affiliation{DAMTP, Centre for Mathematical Sciences, University of
Cambridge, Wilberforce Road, Cambridge CB3 0WA, UK}

\begin{abstract}
We show that if one or more of the `constants' of Nature can vary then
their values, as measured in the laboratory, should oscillate over the year in a very particular way. These seasonal changes in the constants could well be detected, in the near future, with ground-based atomic clocks.
\end{abstract}

\maketitle

Recently, there has been a great deal of interest in the possibility
that some, or all, of the traditional constants of Nature are actually
dynamical and change slowly in space and time, \cite{webb, chand,
  murphyrev, reinhold, localglobal, eorc}.

In this Letter, we show that if the `constants' of Nature do vary,
then their values, as measured by laboratory experiments, should
display and annual variation as the Earth moves around the Sun. We calculate
the magnitude of this effect and find that, although it is expected
to be very small, the continually increasing precision and stability of atomic
frequency standards mean the prospects for detecting it in the near
future are very good.

Theoretical and experimental interest in variation of the constants
has a long history \cite{history}, but the recent renaissance in the
field \cite{localglobal, eorc} has, to a great extent, been motivated by the recent observations of relativistic fine structure in the absorption lines formed in dust clouds around quasars carried out by Webb \emph{et al.}, \cite{webb}.

 Using a data set of 128 objects at redshifts $0.5 < z < 3$, Webb
 \emph{et al.} found the absorption spectra to be consistent with a
 shift in the fine structure constant, $\alpha = e^2/4\pi \epsilon_0
 \hbar c$, between those redshifts and the present day (a period of
 about $10\mathrm{Gyrs}$): $\Delta \alpha/\alpha\equiv (\alpha(z)-\alpha(0))/\alpha(0)=-0.57\pm 0.10\times 10^{-5}$. A smaller study of 23 absorption systems between $0.4\leq
z\leq 2.3$ by Chand \emph{et al.} \cite{chand} found a result consistent with no variation: $\Delta \alpha/\alpha=-0.6\pm 0.6\times 10^{-6}$. However, a recent reanalysis of the \emph{same} data by Murphy \emph{et al.} \cite{murphyrev} was unable to confirm the conclusions of Chand \emph{et al.}, and suggested the revised result: $\Delta \alpha / \alpha =  -0.44 \pm 0.16 \times 10^{-5}$.   Reinhold \emph{et al.} \cite{reinhold} found a $3.5\sigma $
indication of a cosmological variation in another `constant', the proton-electron mass ratio $\mu =m_{p}/m_{e}
$, in their analysis of the vibrational levels of $\mathrm{H}_{2}$ in the absorption spectra
of two quasars at redshifts $z \approx 2.6$ and $z \approx 3.0$; they found $\Delta \mu /\mu =2.0\pm 0.6\times 10^{-5}$
 over the last $12\,\mathrm{Gyrs}$.

Despite these hints of a variation in $\alpha$ and $\mu$,  few would
be prepared to conclude that $\alpha$ and $\mu$ definitely change
with time until either the precision of the astronomical studies is
greatly increased, or a variation in one of these constants can be
directly detected in the more controlled environment of the
laboratory. A significant improvement in the precision of
astronomical studies would most likely require new instrumentation
such as the proposed CODEX spectrograph  \cite{codex}, which is not
expected to be operational before 2017. A firmer
understanding of the potential systematic errors would also be needed. In contrast, the
prospects for a significant improvement, in the near future, in the precision of
laboratory-based varying-constant searches seem much better.

Laboratory constraints on the variation of the constants are
generally found by comparing different atomic frequency standards
over a period of many months or several years.  The most stringent
bound on the temporal variation of $\alpha$ published thus far made use of $6$ years of data and is:
$\dot{\alpha}/\alpha = (-2.6 \pm 3.9) \times 10^{-16} \,
\mathrm{yr}^{-1}$ \cite{peiknew}; if $\alpha$ has varied at
a constant rate over the last $10\,\mathrm{Gyrs}$, then the
findings of Webb \emph{et al.} suggest: $\dot{\alpha}/{\alpha} =
(6.4 \pm 1.4) \times 10^{-16}\,\mathrm{yr}^{-1}$ \cite{webb}.  An
important motivation for this Letter is that the ability of
laboratory tests to measure changes in $\alpha$ seems likely to improve
markedly in the near future.  The ACES (Atomic Clock Ensemble in
Space) project, currently projected to fly on the International
Space Station in 2010, will be able to constrain $\dot{\alpha}/\alpha$ at the
$10^{-17}\,\mathrm{yr}^{-1}$ level. Recently, Cing\"{o}z \emph{et al.} \cite{cingoz}
reported a new limit on $\dot{\alpha}/\alpha$ found by monitoring
the transition frequencies between two nearly degenerate,
opposite-parity levels in two isotopes of atomic Dysprosium (Dy)
over 8 months.  These energy levels are particularly
sensitive to changes in $\alpha$ \cite{flamth}. Cing\"{o}z \emph{et
al.} found that $\dot{\alpha}/{\alpha} = -(2.7 \pm 2.6)\times
10^{-15}\,\mathrm{yr}^{-1}$, but importantly they estimate that an
ultimate sensitivity to changes, $\delta \alpha$, in $\alpha$ of one
part in $10^{18}$ is feasible. Moreover, Flambaum  \cite{flamth} has recently noted that an even greater
improvement in precision could be achieved by making use of the
enhanced effect of $\alpha$ variation on the very narrow UV
transition  between the ground and first excited state of the
${}^{299}$Th nucleus.  The corresponding experiment
could potentially detect a non-zero $\vert \delta{\alpha}/\alpha \vert$ as small
as $10^{-23}$; 7 orders better than current bounds. Despite the expected increase in the precision of
laboratory tests, any such experiment must still run for many months, or even
several years, if it is to place tight constraints on any
time variation. In this Letter we note that if a
`constant', $\mathcal{C}$, can vary, then its value, as measured in the laboratory, should vary during the
year.  This variation will have a very distinctive shape and be
correlated with the Earth's distance from the Sun.  These two
properties should make it easier to separate any signal from noise. Additionally,  an experiment would only need to run for six months to constrain any fluctuation in the constants due to this effect.

Variation of some, or all, of the constants of Nature is fairly generic
prediction of most modern proposals for fundamental physics beyond the standard model. Indeed, it is one of the few low-energy signatures of such theories. At the low energies appropriate to classical physics, the values of the constants are determined by the vacuum expectation of a scalar field, or \emph{dilaton}, $\phi$ and $\alpha = \alpha(\phi)$, $\mu = \mu(\phi)$. Henceforth, we use units in which $c=\hbar = 1$. It is generally assumed that the scalar field theory associated with $%
\phi $ has a canonical kinetic structure, and that variations in
$\phi$ conserve energy and momentum, as well as contributing to the
curvature of space-time in the usual way.  These considerations imply
that $\phi$ satisfies a non-linear wave equation \cite{localglobal}:
\begin{equation}
\square \phi =\frac{1}{\omega}\left(\sum_{i} \frac{\partial \mathcal{C}_{i}(\phi)}{\partial \phi} \frac{\delta \mathcal{L}_{m}}{\delta \mathcal{C}_{i}} + \frac{\partial V(\phi)}{\partial \phi}\right),
\label{gen}
\end{equation}
where $\mathcal{L}_{m}$ is the Lagrangian density for the matter
fields, and the $\mathcal{C}_{i}(\phi)$ represent different,
$\phi$-dependent, `constants' of Nature; $V(\phi)$ is some
self-interaction potential for the dilaton,  and $\omega$ is a
constant with units of $(mass)^2$ which sets the strength with which
$\phi$ couples to matter.  General expectations from string theory
suggest that $4\pi G \omega \sim O(1)$ and that if a constant, $\mathcal{C}_{i}$,
is dynamical then $\partial \ln \mathcal{C}_{i} / \partial \phi \sim
O(1)$.  Ultimately, it is our goal to determine, or bound, the
parameters of such theories experimentally. Solar-system tests of
gravity currently seem to prefer $4\pi G \omega \gg 1$ \cite{cwill}.
A similar equation to Eq. (\ref{gen}) applies when
variations in the constants are driven by multiple scalar fields.
Multiple fields might complicate the cosmological evolution of the
`constants' but, over solar system scales, the dynamics of each
field are well-approximated by a wave equation with the form of Eq.
(\ref{gen}). At the present time, the matter to which $\phi$ couples
is non-relativistic.  Defining $\rho_{j}$ to be the
energy-density of the $j^{\mathrm{th}}$ matter species, we then have $\mathcal{L}_{m} \approx \sum_{j}\rho_{j}$ and Eq. (\ref{gen})
reduces to:
\begin{equation}
\square \phi =\frac{1}{\omega}\left(2\sum_{j} \zeta_{j}(\phi) \rho_{j} + \frac{\partial V(\phi)}{\partial \phi}\right)= \frac{2\zeta(\phi) \rho}{\omega} + \frac{V_{,\phi}}{\omega},
\label{gen2}
\end{equation}
where $2\zeta_{j}(\phi) = \sum_{i} \mathcal{C}_{i,\phi} \delta (\ln
\rho_{j}) / \delta C_{i}$, $\rho = \sum_{j} \rho_{j}$ is the total
energy density of matter, and $\zeta(\phi) = \sum_j \zeta_j
\rho_j/\rho$. The value of the $\zeta$ for a body, or a system of
bodies, generally depends on its composition.  The effective
mass-squared of the scalar field is given by $m_{\phi}^2=V_{,\phi
  \phi}/\omega$. Generally, if time variations in a constant occur at
anything approaching a detectable level then $m_{\phi} \lesssim
10^{-63}\mathrm{g} \sim H_0$, where $H_0$ is the Hubble parameter today.   If $V(\phi)$ is highly non-linear, and $\zeta/\omega$ and $V_{,\phi}$ have opposite signs, then $\phi$ may behave as a \emph{chameleon field} (see refs. \cite{cham}). The mass of $\phi$ would then depend heavily on $\rho$. Whilst chameleon field theories are very interesting, late-time variation of the fundamental constants is negligible in all known, experimentally viable,  chameleon models. For this reason, we do not consider such theories here, and henceforth assume that $\phi$ is \emph{not} a chameleon field.

Tests of gravity constrain any variations in $\phi$ in the solar
system to be very small \cite{cwill}. We can therefore linearise Eq.
(\ref{gen2}) about the background value of the field, $\phi_b(t)$,
which will track the cosmological value of $\phi$, \cite{localglobal},
i.e. we take $\phi \approx \phi_b(t) + \delta \phi$. The majority of recent
laboratory-based searches for varying-constants have looked for changes in $\alpha$.  We therefore focus our attention on theories in which $\alpha$ varies, and scale $\phi$ and $\omega$
so that $\partial \ln \alpha(\phi_b)/ \partial \phi = 2$ today.
Although we primarily consider the annual variations of $\alpha$,
similar variations should be expected in all varying constants.

\begin{figure}[tbh]
\begin{center}
\includegraphics[scale=0.47]{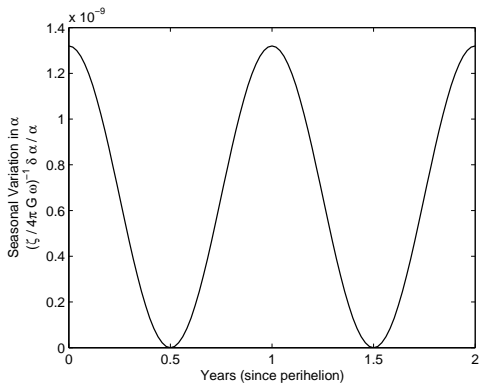}
\end{center}
\caption{The shape of the expected annual variations in $\alpha$
over 2 years.  We have defined $\delta \alpha /\alpha = (\alpha -
\alpha_{ap})/\alpha_{ap}$, where $\alpha_{ap}$ is the value of
$\alpha$ at aphelion.  The magnitude of the fluctuations is
proportional to $\zeta_{\odot}/4\pi G \omega$.} \label{fig1}
\end{figure}

In the solar system, any temporal-gradients in $\delta \phi$ are
expected to be small compared to the spatial ones (see
Ref. \cite{localglobal}) and so, at leading order, we may replace the
wave-operator, $\square$, by the Laplacian, $\nabla^2$, giving:
\begin{equation}
\nabla^2 \delta \phi = \frac{2\zeta}{\omega} \rho + m_\phi^2 \delta \phi,
\end{equation}
where $m_{\phi}$ and $\zeta$ independent of $\delta \phi$, although
$\zeta$ is still composition dependent. Solving this equation, we find
that the Sun induces the following contribution to $\delta \phi$:
\begin{equation}
\delta \phi_{\odot} = \frac{\zeta_{\odot}}{4\pi G \omega}
\frac{2GM_{\odot}}{r}e^{-m_\phi r}, \label{phisun}
\end{equation}
where $r$ is the distance to the Sun's centre of mass, $M_{\odot}$
is the Sun's mass, and $\zeta_{\odot}$ is the value of $\zeta$ for
the Sun.   The Earth's orbit around the Sun is \emph{not} perfectly circular and so the distance of the
Earth from the Sun changes slightly over the year, fluctuating by about 3\% from
aphelion to perihelion.  As a result, the value of $\delta \phi_{\odot}$, and hence also
$\alpha$, will oscillate annually as $r$ changes.  The values of any
other constants of Nature that depend on $\phi$ will also vary
throughout the year. Experimental searches for a time-variation in $\alpha$
generally assume an approximately linear drift in the $\alpha$;
however, these seasonal changes are clearly
oscillatory. Attempting to fit a linear drift to an annual oscillation would lead to inaccurate conclusions.  The change in $\alpha$ from aphelion to
perihelion due this annual fluctuation is given by $\delta \alpha / \alpha \approx 2(\delta \phi_{\odot}(r_{ap})
- \delta \phi_{\odot}(r_{per}))$.  If $m_{\phi}(r_{ap}-r_{per})/2 \ll
1$, we predict:
\begin{eqnarray}
\left.\frac{\delta \alpha}{\alpha}\right\vert_{seasonal} =  -1.32 \times 10^{-9}\frac{\zeta_{\odot}}{4\pi G\omega}e^{-m_{\phi}a},
\end{eqnarray}
where $a= 1.496 \times 10^{8}\,\mathrm{km}$ is the radius of the
Earth's semi-major axis. If measurements are taken throughout the
year then seasonal changes in $\alpha$ should be distinguishable
from noise, or any linear drift, by their distinctive shape, which is shown in
FIG. \ref{fig1}.  It is also important to note that, whereas cosmological changes in the constants will virtually non-existent if $m_{\phi} \gg
10^{-63}\,\mathrm{g} \sim H_{0}$, the seasonal fluctuations noted
here only require  $m_{\phi}\lesssim 10^{-51}\,\mathrm{g} \sim
1\,\mathrm{AU}^{-1}$ to be potentially detectable.

Current laboratory searches for a variation $\alpha$ find a $\delta \alpha /\alpha$ consistent with no variation at the $2 \times 10^{-15}$ level \cite{peiknew,cingoz}.  This corresponds to a bound on $\zeta_{\odot}/4\pi G \omega$ of:
$$
\left\vert\frac{\zeta_{\odot}}{4\pi G \omega}\right\vert < 2 \times 10^{-6}.
$$
The quantity $\zeta_{\odot}/4\pi G \omega$
also sets the magnitude of violations of local position invariance
(LPI) \cite{eorc}.  Currently, the best bounds on the validity of
LPI only give a limit of $\vert \zeta_{\odot} / 4\pi G \omega\vert < 10^{-4}$ \cite{cwill}; two
orders of magnitude \emph{worse} than is already achievable by making use
of the Earth's motion around the Sun.  If $\delta \alpha/\alpha$ can
be measured at the $10^{-18}$ level then this would correspond to a
measure of $\zeta_{\odot}/4\pi G \omega$, and any violation of LPI, at
the $10^{-9}$ level.  Similarly, if a
${}^{299}\mathrm{Th}$-based experiment can
be carried out, and $\delta \alpha /\alpha$ constrained to within
one part in $10^{23}$,  $\zeta_{\odot}/4\pi G \omega$ and the
validity of LPI can be measured to within one part in $10^{14}$.
Schiller \emph{et al.} \cite{schiller} recently noted that space-based atomic clocks
could utilise the altitude dependence of the Earth's gravitational
potential to provide a similar increase in the precision to which violations of
LPI can be measured.  Whilst this is an excellent
idea, the potential improvement in precision they find is no better than
can be achieved, much more cheaply, on the ground by making use of
the eccentricity of the Earth's orbit.

In general, the $\zeta$ parameter is composition dependent and, as a
result, the weak equivalence principle (WEP) is violated at some level.
The magnitude of the expected WEP violations over distance of
$1\,\mathrm{AU}$ will also depend on $\zeta_{\odot}/4\pi G \omega$.
We now use bounds on any WEP violation to estimate the precision to which
$\delta \alpha /\alpha$ must be measured before any annual
fluctuations in $\alpha$ could potentially be detected.    Over distance of about
$1\mathrm{AU}$, the best current constraint on WEP violations was
reported in Ref. \cite{RotWash}. Using a modified torsion balance, the differential acceleration
towards the Sun of two test bodies was measured and found to be:
$$
\eta = \frac{2\vert a_{1} - a_{2} \vert}{\vert a_1 + a_2 \vert} = 0.1 \pm 2.7 \pm 1.7 \times 10^{-13},
$$
which is consistent with no violation of WEP; here $a_1$ and $a_2$ are respectively the accelerations of the first and second test body towards the Sun. The E\"{o}tvos parameter, $\eta$, measures the strength of WEP violations. The reasonably general varying-constant model used throughout this Letter predicts:
\begin{equation}
\eta = 4\vert \zeta_{1} - \zeta_{2} \vert \left\vert\frac{\zeta_{\odot}}{4\pi G \omega}\right\vert.
\end{equation}
In some varying-constant theories, such as the varying-speed of
light (VSL) model considered in Ref. \cite{eorc}, $\zeta_1 =
\zeta_2$ for all bodies and so WEP is \emph{not} violated.  For such
theories, WEP violation searches \emph{cannot} constrain
$\zeta_{\odot}/4\pi G \omega$.   It is more generally the case, however, that
WEP \emph{is} violated in theories where some or all `constants' can vary, and that $\vert\zeta_1 - \zeta_2\vert \sim O(10^{-1}\vert
\zeta_{\odot} \vert)$ for such experiments. For varying-$\alpha$
theories, we additionally expect $10^{-4} \lesssim \vert \zeta \vert
\lesssim 1$ for baryonic matter.  The smallest values of $\zeta$ are expected for
theories in which $\alpha$ is the only $\phi$-dependent constant.
If $\Lambda_{QCD}$ is also $\phi$-dependent then generally $\zeta \sim
O(1)$. Therefore, in a scalar field theory,
which describes a variation in $\alpha$ and violates WEP, we expect:
$$
\left\vert \frac{\delta \alpha}{\alpha} \right\vert_{seasonal} \lesssim 10^{-21}-10^{-17},
$$
where larger values of $\vert \delta \alpha /\alpha \vert$ are more feasible if $\zeta$
is smaller. If, as seems likely, $\delta \alpha /\alpha$ can measured to a precision of one
part in $10^{18}$ in the near future, then constraints on annual oscillations of $\alpha$
could well be detected or constraints on some
varying-$\alpha$ theories improved by at least an order of
magnitude.  If $\delta \alpha / \alpha$ can be constrained at the
$10^{-23}$ level, as suggested by Flambaum \cite{flamth}, then the
prospects for detecting the predicted seasonal variation in $\alpha$
are even better, and, in almost all cases, direct laboratory
bounds on $\delta\alpha/\alpha$ would more tightly constrain
varying-$\alpha$ theories than WEP violation tests currently
do. Indeed, if the $10^{-23}$ precision can be reached, then we could
derive a model-dependent constraint on WEP violations due to
varying-$\alpha$ at the one part in $10^{14}-10^{19}$ level; the
greatest precision is for theories  with small $\zeta$.  For comparison,
proposed satellite-based EP experiments such as MICROSCOPE, GG and
STEP promise a model-independent measure of any WEP violation at the 1
part in $10^{15}$, $10^{17}$ and $10^{18}$ levels respectively \cite{newsats}. 

WEP violation searches are sensitive only to the magnitude of the
$\zeta$ parameters and not to their sign.   As noted by Magueijo,
Barrow and Sandvik in Ref. \cite{eorc}, knowledge of the sign of $\zeta$ for
baryonic matter is very important when attempting to discern between
two different varying-$\alpha$ models.  Varying-$\alpha$
theories that are most naturally interpreted as a change in the
fundamental electric charge $e$, $\zeta > 0$ is natural for baryonic
matter, \cite{eorc}. For such theories, the observations of Webb \emph{et
  al.}, \cite{webb}, suggest that either $\zeta < 0$ for dark matter,
or that the cosmological evolution of $\alpha$ is dominated by the
potential ($V_{,\phi}/\omega$) term in Eq. \ref{gen2} \cite{eorc}. In contrast, models that are simplest when viewed
as describing a varying speed of light predict $\zeta < 0$.  Unlike
WEP violation searches, the annual fluctuations in $\alpha$ are
sensitive to the sign of $\zeta$. If $\alpha$ does vary, and the
seasonal variation could be detected, then sign of
$\zeta_{\odot}$ would be known.  Additionally, when combined with
WEP violation bounds, such a measurement would also allow the values
of both $\zeta_{\odot}$ and $\omega$ to be deduced separately.
Achieving this would greatly increase our understanding of the
theory that underpins any variation in $\alpha$.

We have noted, in this Letter, that if some of the constants of Nature vary,
then we should expect their values on Earth to oscillate seasonally.
The estimated upper bound on the magnitude of these fluctuations is
such that they could well be detected in the near future by atomic-clocks. Importantly, the search for these variations
would not require the creation of a totally new experimental
programme, as laboratory searches for a temporal drift in the
constants would also be sensitive to these yearly oscillations.  It
is feasible that such experiments could measure variations in
$\alpha$ at the $10^{-18}$ level \cite{cingoz}, and there have
been suggestions that changes as small as one part in the $10^{23}$
could be detected \cite{flamth}.   It is also important to note that, in many
theories, the expected magnitude of the annual oscillations in a
`constant', such as $\alpha$, will be comparable to, if not much larger
than, the expected yearly drift in the value of the constant.   Indeed, if
$10^{-63}\,\mathrm{g} \gg  m_{\phi} \lesssim 10^{-51}\,\mathrm{g}$,
then any linear temporal drift in the `constant' would be imperceptibly small,
whereas the magnitude of the annual variations would be at the
$10^{-9} \zeta_{\odot}/4\pi G \omega$ level.  Furthermore, even if
$\phi$ is very light ($m_{\phi} \lesssim 10^{-63}\,\mathrm{g}$), the
cosmological evolution of $\phi$ is, in some theories, primarily
driven by the coupling of $\phi$ to matter (i.e. the effect of
the $V_{,\phi}/\omega$ in Eq. \ref{gen2} is negligible).  Such
models predict that today $\dot{\alpha}/\alpha \approx -(3H^2_0 t_{U})/(2\pi G \omega)\left(\zeta_{dm}\Omega_{dm} + \zeta_{b}\Omega_{b}\right)$,
where $\Omega_{dm}$ and $\Omega_{b}$ are respectively the density of
the dark and baryonic matter in the Universe today as a fraction of
the critical density, $t_U$ is the age of the Universe, and
$\zeta_{dm}$ and $\zeta_{b}$ are the respective values of $\zeta$ for
dark and baryonic matter. Using values for $\Omega_{dm}$, $\Omega_{b}$, $t_U$ and $H_0$
from WMAP, \cite{spergel}, we see that, over the course of a year,
$\alpha$ is expected to change by a fractional amount:
$$
\left.\frac{\delta {\alpha}}{\alpha}\right\vert_{cosmo} \approx -\frac{(9.7\pm 0.9)\zeta_{dm} + (1.9 \pm 0.1) \zeta_b}{4\pi G \omega} \times 10^{-11}.
$$
We expect $\zeta_{b} \sim O(\zeta_{\odot})$ and so, unless
$\zeta_{dm} \gtrsim 13 \zeta_{\odot}$, the magnitude of the annual
changes in $\alpha$, identified in this Letter, will be greater
than any linear temporal drift in its value.  Failure to allow for these seasonal changes, could serious compromise the analysis of data found by
laboratory searches for variations in  $\alpha$. 

In this Letter we
have shown that if one or more of the `constants' of Nature can
vary, then the non-zero eccentricity of the Earth's orbit will cause
their values, as measured in the laboratory, to vary annually in a very
particular way.  Recent and expected advances in the precision and stability of frequency standards make it feasible that such a seasonal change in $\alpha$
could be detected, without the need to perform space-based tests, in
the near future.  Improved constraints on any such oscillation will
greatly enhance our understanding of varying-constant theories and
would also be used to improve current bounds on local
position invariance. By both these means, we will be able to constrain
constrain the nature of fundamental physics beyond the standard model.

\acknowledgments This work was supported by PPARC.  I would
like to thank J. D. Barrow for reading a preprint of this Letter and for  helpful comments and suggestions.


\begin{thebibliography}{99}

\bibitem{webb} J.~K.~Webb \emph{et al.}, Phys. Rev. Lett. \textbf{82}, 884 (1999); \textbf{87}, 091301 (2001).

\bibitem{chand} H.~Chand \emph{et al.}, Astron. Astrophys. \textbf{417}, 853
(2004); R.~Srianand \emph{et al.},, Phys. Rev. Lett. \textbf{92}, 121302 (2004).

\bibitem{murphyrev} M.~T.~Murphy \emph{et al.}, \texttt{astro-ph/0612407}.

\bibitem{reinhold} E.~Reinhold \emph{et al.}, Phys. Rev. Lett. \textbf{96}, 151101
(2006).

\bibitem{localglobal} D.~J.~Shaw and J.~D.~Barrow, Phys. Rev. D
\textbf{73}, 123505 (2006); \textbf{73}, 123506 (2006);
Phys. Lett. \textbf{B639} 596-599 (2006).

\bibitem{eorc} J. Magueijo, J. D. Barrow and H. B. Sandvik,
  Phys. Lett. \textbf{B549}, 284-289 (2002); H. B. Sandvik, J. D. Barrow, J. Magueijo,
  Phys. Rev. Lett. \textbf{88}, 031302 (2002).

\bibitem{history} P. A. M. Dirac, Nature \textbf{139}, 323 (1937);
  G. Gamow, Phys. Rev. Lett. \textbf{19}, 757 (1967); \textbf{19}, 913
  (1967).

\bibitem{codex} L.~Pasquini \emph{et al.}, in \emph{Scientific Requirements for Extremely Large Telescopes}, proceedings of the 232nd Symposium of the IAU, Cape Town, South Africa, edited by P. A. Whitelock, M. Dennefeld and B. Leibundgut (CUP, Cambridge, 2006); P.~Molaro \emph{et al.}, \emph{ibid}.

\bibitem{peiknew} E.~Peik \emph{et al.}, to appear in
  \emph{Proceedings of the 11th Marcel Grossmann Meeting}, edited by
  H. Kleinert, R.T. Jantzen and R. Ruffini (World Scientific, Singapore, 2007).

\bibitem{cingoz} A.~Cing\"{o}z \emph{et al.},
  Phys. Rev. Lett. \textbf{98}, 040801 (2007).

\bibitem{flamth} V.~V.~Flambaum, Phys. Rev. A \textbf{73}, 034101 (2006).

\bibitem{cwill} C.~M.~Will, Living Rev. Relativity \textbf{9}, 3 (2006).

\bibitem{cham} J.~Khoury and A.~Weltman, Phys. Rev. Lett. \textbf{93}, 171104 (2004); D. F. Mota and D. J. Shaw, \emph{ibid.} \textbf{97}, 151102 (2006).



\bibitem{schiller} S.~Schiller \emph{et al.}, in \emph{Proceedings of
  the III\grad  International Conference on Particle and Fundamental
  Physics in Space, Beijing, 2006}. \texttt{gr-qc/0608081}.

\bibitem{RotWash} S.~Baessler {\it et al.}, Phys. Rev. Lett. {\bf 83}, 3585 (1999).



 \bibitem{newsats} P.~Touboul \emph{et
  al.}, Acta Astronaut. \textbf{50}, 433 (2002); A.~M.~Nobili \emph{et
  al.}, Class. Quant. Grav. \textbf{17}, 2347 (2000); J.~Mester
  \emph{et al.}, \emph{ibid.} \textbf{18}, 2475 (2000).

\bibitem{spergel} D.~N.~Spergel \emph{et al.}, \texttt{astro-ph/0603449}.

\end{thebibliography}
\end{document}